\documentclass[twocolumn,showpacs,preprintnumbers,amsmath,amssymb]{revtex4}

\usepackage{graphicx}
\usepackage{dcolumn}
\usepackage{bm}
\usepackage{epsfig}
\usepackage{graphics}
\usepackage{graphicx}
\usepackage{mathtext}

\begin{document}

\title{
Self-similarity, Aboav-Weaire's and Lewis' laws in weighted planar stochastic lattice
}%

\author{F. R. Dayeen and M. K. Hassan
}%
\date{\today}%

\affiliation{
 Department of Physics, University of Dhaka, Dhaka 1000, Bangladesh
}

\begin{abstract}%

In this article, we show that the block size distribution function in the weighted planar stochastic lattice (WPSL), 
which is a multifractal and whose dual is a scale-free network, exhibits dynamic scaling.
We verify it numerically using the idea of data-collapse. As the WPSL is a space-filling cellular structure, we thought
it was worth checking if the Lewis and the  Aboav-Weaire laws are obeyed in the WPSL. To this end, we find that the 
mean area $<A>_k$ of blocks with $k$ neighbours grow linearly up to $k=8$, and hence the Lewis law is obeyed. 
However, beyond $k>8$ we find that $<A>_k$ grows exponentially to a
constant value violating the Lewis law. On the other hand, we show that the Aboav-Weaire law is violated 
for the entire range of $k$. Instead, we find that the mean number of neighbours of a block adjacent to a block with
 $k$ neighbours is approximately equal to six, independent of $k$. 

\end{abstract}


\maketitle

\section{Introduction}


Space-filling planar cellular structures are found in a wide variety of seemingly disparate physical and biological systems.
 Examples include grain structures in polycrystals, 
cell texture and tissues in biology, acicular texture in martensite growth, tessellated pavement on ocean shores, soap froths
and agricultural land division according to ownership etc. just to name a few \cite{ref.polycrystal, ref.biocell, ref.soapfroths}. 
The question of how these structures appear  and the thirst for understanding  their topological and geometrical properties  
have always attracted interest among scientists in general and physicists in particular.
To this end, there exists a number of models that prescribe how to generate cellular structures. Either these
structures themselves or their properties can mimic structures found in nature. 
In general, cellular structures appear through random tessellation, tiling, or 
subdivision of a plane into contiguous and nonoverlapping cells. For instance, Voronoi lattice and Apollonian packing are 
formed by partitioning or tiling of a plane into contiguous and non-overlapping convex polygons and 
disks respectively \cite{ref.model,ref.apollonian}. 

Two empirical laws, namely the Lewis and the Aboav-Weaire laws, have been found to play a key role in studies of planar
cellular systems. The two laws characterize the two most prominent properties of the cellular structure.
For instance, the Aboav-Weaire law is about
 nearest-neighbor correlations of the cells in the structure \cite{ref.Aboav_0, ref.Weaire_0}. It states that 
the average number of neighbors $m_n$ of a typical cell that neighbors 
an $n$-sided cell obeys the following relation
\begin{equation}
\label{eq:aboav}
m_n=a+{{b}\over{n}}, 
\end{equation}
where $a$ and $b$ are constant. It implies that many-sided cells tend to have few-sided neighbors and vice versa.
Recent experiments and
numerical simulations suggest that the Aboav-Weaire law plays a significant role during grain growth 
\cite{ref.grain_growth_0, ref.grain_growth_1,ref.grain_growth_2}. There have already been many attempts
to justify the emperical and simulation data theoretically \cite{ref.grain_growth_3}. 

The Lewis law on the other hand is yet another relation which states that the normalized mean area $<A>_n$ of an arbitrarily
chosen $n$-sided cell increases with $n$ as
\begin{equation}
\label{eq:lewis_law}
{{<A>_n}\over{<A>}}=\alpha n-\beta,
\end{equation}
where $<A>$ is the mean area of all cells and $\alpha$, $\beta$ are constant. Lewis observed that in several
2D cellular structure Eq. (\ref{eq:lewis_law}) is obeyed at least up to a certain value
of $n$ at various stages of their growth \cite{ref.Lewis_0, ref.Lewis_1,ref.Lewis_2}. For
Poisson Voronoi tessellations, it has been shown emperically that $<A_n>$ varies
linearly with $n$ for $n<11$ \cite{ref.Lewis_3, ref.Lewis_4, ref.Lewis_5}. Flyvbjerg proposed a nonlinear dynamic
model through which he was able to derive Lewis' law which is asymptomatically valid 
\cite{ref.Flyvberg}. Planar cellular structures roughly have two broad classes
of applications. Firstly, they may directly model cellular structures occuring in great many different situations such as
biological tissue or soap froths etc. Secondly, they may
serve as a skeleton on which one can study his favorite theory or model on it such as percolation and various spreading
phenomena. The later case, is interesting
when the structure is topologically disordered yet possesses properties which are time and size independent.

The two empirical laws in question  have mostly been 
tested to the structures which have two properties. Firstly, cellular structures where it is almost impossbile to find cells  
that have significantly higher or fewer neighbours than the average. That is, there exists a characteristic 
value so that the coordination number distribution is peaked around a mean value. 
Secondly, cellular structures in which none of the cells have any side which is shared among sides of more than one cell. 
That is, a cell which has $m$ sides can have no more or no 
less than exactly the same number of nearest neighbors. In nature, planar cellular systems come in a wide variety.
For instance, we may have a planar cellular structure where the coordination number distribution may not be peaked around 
a typical or mean value. Instead,
it may vary over many orders of magnitude, or may even have a distribution function that follows power-law. 
Besides, two cells can share only a portion of a side instead of the whole side. That is, 
the number of neighbors of a cell can be higher than the number of sides a given cell has. 
Moreover, cellular structure may emerge through evolution where cells can be of different sizes and have a 
great many different number of neighbors since nature favours these properties as a matter of rule rather than exception. 

Recently, one of us Hassan {\it et al} proposed a space-filling weighted planar
stochastic lattice (WPSL) where the coordination number distribution function follows power-law \cite{ref.wpsl_njp,ref.wpsl_jphys}. 
This is in sharp contrast to many of the cellular structures that we are familiar with.
In this article, we first investigate the self-similar properties of the WPSL. To this end, we show
that the block area size distribution function exhibits dynamic scaling which we proved using the idea of data-collapse.
We then investigate and find whether the Lewis and Aboav-Weaire laws are obeyed in the WPSL. It is
an interesting proposition since the two laws have never been chacked in a scale-free cellular planar structure which is also
a multifractal. We find that the normalized mean area $<A>_n$ of an $n$-sided cell does grow neither linearly
nor indefinitely as stated by the Lewis law over the entire range of $n$ values. Instead, we find that it 
attains its maximum value and it grows to this maximum value exponentially. In fact, the linear growth of the
 normalized mean area is found true only for $n$.  On the other hand, we find that the Aboav-Weaire law is obeyed
for the entire range of $n$ values. Encouraged by this, we also check if the Barabasi-Albert networks, 
the best-known thoretical model for scale-free network, and find that it too obeys the Aboav-Weaire law.


The rest of the article is organized as per the following scheme. In section II we briefly define our model.
In section III we show that the area size distribution function of the WPSL exhibits dynamic
scaling and we prove it using the idea of data-collapse. Thereafter in section IV we present our findings about 
two laws namely the Lewis and the Aboav-Weaire laws as regard to WPSL. Finally in section V
we summerize our findings and suggest possible future directions for further work.

\section{Definition of the WPSL}

\begin{figure}
\includegraphics[width=8.5cm,height=8.0cm,clip=true]{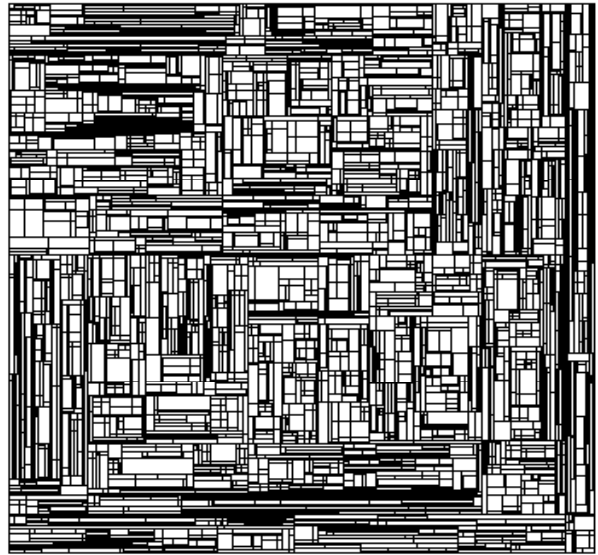}
\caption{A snapshot of the weighted stochastic lattice.
}
\label{fig1}
\end{figure}%

To appreciate the model better, we find it worthwhile first to describe 
the construction process of the square lattice.
It can be best understood by choosing an initiator, say a square of unit area, and by
defining a generator that divides it into four equal squares or blocks. In the next step and steps thereafter the generator is applied 
recursively to all the available blocks. As this process continues it will eventually generate
a square lattice. Now the interesting questions to follow are: What if we define a generator that divides the initiator randomly 
into four blocks instead of four equal blocks? What if, in step two and thereafter the generator is applied
to only one of the available blocks at each step by picking it preferentially with respect to the areas? 
That is, in step one, the generator divides the initiator 
randomly into four smaller blocks. In step two, one of the four new blocks is picked with probability equal
to their respective area, and the generator is applied to divide it randomly into four blocks. In step three, we again
pick one of the seven blocks prferentially with respect to their area and apply the generator. The idea about the future steps
is perhaps made amply clear enough. As the process continues for a sufficiently long time, where each step is defined as
one time unit, it will result in a cellular structure which we call weighted planar stochastic lattice. 
A snapshot of the WPSL at late stage (figure 1) provides an awe-inspiring perspective on 
the emergence of an intriguing and rich pattern of blocks. (see Fig. (\ref{fig1})).

Perhaps, giving an exact algorithm can define the model better than the mere definition.
In general, the $j$th step of the algorithm can be described as follows.
(i) Subdivide the interval $[0,1]$ into $(3j-2)$ subintervals of size $[0,a_1]$, $[a_1, a_1+a_2]$,$\  ...$, 
$[\sum_{i=1}^{3j-3} a_i,1]$ each of which represents the blocks labelled by their areas $a_1,a_2,...,a_{(3j-2)}$ respectively.
(ii) Generate a random number $R$ from the interval $[0,1]$ and find which of the $(3i-2)$ sub-interval 
contains this $R$. The corresponding block it represents, say the $p$th block with area $a_p$, is picked.  
(iii) Calculate the length $x_p$ and the width $y_p$ of this block and keep note of the coordinate of the
lower-left corner of the $p$th block, say it is $(x_{low}, y_{low})$.
(iv) Generate two random numbers $x_R$ and $y_R$ from $[0,x_p]$ and $[0,y_p]$ respectively and hence
the point $(x_{R}+x_{low},y_{R}+y_{low})$ mimics a random point chosen in the block $p$.
(v) Draw two perpendicular lines through the point $(x_{R}+x_{low},y_{R}+y_{low})$ 
parallel to the sides of the $p$th block in order to divide it into four smaller blocks. The label $a_p$ is now redundant and hence
it can be reused.
(vi) Label the four newly created blocks according to their areas $a_p$, $a_{(3j-1)}$, $a_{3j}$ and $a_{(3j+1)}$ respectively 
in a clockwise fashion starting from the upper left corner.
(vii) Increase time by one unit and repeat the steps (i) - (vi) {\it ad infinitum}.

\section{Area size distribution function and dynamic scaling}

The snapshot of the WPSL shown in Fig. (\ref{fig1}) provides a clear impression of how its structure should look like
in the long time limit. Clearly, it looks seemingly complex, manisfestly intricate 
and inextricably intertwined, which makes it an interesting candidate to look
deep into and check if there is some order however disordered it looks. Such a lattice that emerges through evolution can only be useful if 
the snapshots taken at different late stages are similar. It makes various characteristic properties of the lattice independent of its size. In physics, similarity has a specific meaning.  
Two snapshots of WPSL taken at two very different times can be similar if one of its governed quantity, say $f(x,t)$, exhibits
dynamic scaling. The function $f(x,t)$ is said to obey dynamic scaling if it has the form
\begin{equation}
\label{eq:3}
f(x,t)\sim t^\theta \phi(x/t^z),
\end{equation}
where exponents $\theta$ and $z$ are fixed by the dimensional relations $[t^\theta]=[f]$ and $[t^z]=[x]$ respectively, 
while $\phi(\xi)$ is known as the scaling function \cite{ref.family_Vicsek}. Finding dynamic scaling in any system has always 
represented progress for researchers as it implies that the phenomena 
that it represents is self-similar. One of us found such self-similarity 
in many different processes like the kinetics of aggregation, stochastic Cantor set and in complex network theory  
\cite{ref.aggregation,ref.cantor,ref.badc}.

\begin{figure}
\includegraphics[width=8.50cm,height=5.5cm,clip=true]{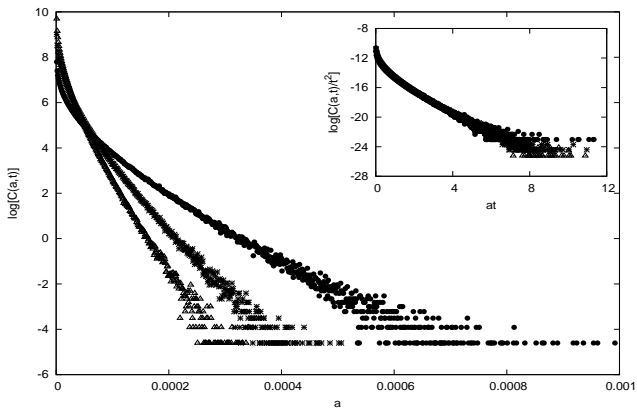}
\caption{Log-linear plot of the area size distribution function $C(a,t)$ vs $a$ of the WPSL 
 for three different times $t=10,000,20,000,30,000$. 
\label{fig2}
 }
\end{figure}

One of the interesting observable physical quantities for the WPSL can well be the block size or area distribution function $C(a,t)$. 
We define it  
such that $C(a,t)da$ describes the concentration
of blocks of area within the size range $a$ and $a+da$ at time $t$. We
 find it worthwhile to check if $C(a,t)$ exhibits self-similarity or not. 
To get a sense of how the distribution function $C(a,t)$ varies with area $a$ at different fixed time $t$, we collect data at
three different instants, say at $t_1=10,000, t_2=20,000, t_3=30,000$. We can use the resulting data for each
of the three different times to plot a histogram that describes the occurrence frequency or 
the number of blocks within a given class. However, we find it convenient to normalize the occurrence frequency
 by the width $\Delta a$ of the interval size so that area
under the histogram 
\begin{equation}
\label{eq:conservation_law}
 \int_0^\infty aC(a,t)da=1,
\end{equation}
gives the area of the initiator.
The histogram thus represent $C(a,t)$ as a function of $a$ for fixed time.  However, the difference among the plots for 
$C(a,t)$ at different times can be best appreciated if we plot $C(a,t)$ versus $a$ in 
the $\log$-linear scale which is shown in Fig. (\ref{fig2}). It is clearly linear, at least  near the tail, revealing that 
$C(a,t)$ decays exponentially but only for large $a$. 
In order to find a better form and interpretation for $C(a,t)$ we invoke the idea of dimensional analysis below.



Note that the sum of the areas $A$ of all the blocks $N(t)$ present at any given time $t$ is $\sum_{i=1}^N a_i=1$  since
we choose a square of unit area as the initiator. The number of blocks $N(t)$ at time $t$ on the other hand is $N(t)=1+3t$.
The mean area $<a(t)>=A/N(t)$ therefore scales as $<a(t)>\sim t^{-1}$. It implies that one of the two governing
parameters $a$ and $t$ of $C(a,t)$ can be chosen as an independent parameter. Let us choose time $t$ be the independent
quantity so that we can express $C$, which is physical quantity, in terms of $t$ alone i.e., $C=C(t)$. Note
that the dimension function of a physical quantity always obeys power-monomial law and hence we can write  $\sim t^\theta$.
We therefore can define a dimensionless governing parameter $\xi=at$
and a dimensionless governed parameter $\Pi=C(a,t)/t^{\theta}$.
Note that the numerical value of $\Pi$ should remain the same even if the unit of 
time $t$ is changed by some factor because it is a dimensionless quantity. 
However, the numerical value of $\Pi$ still may depend on the dimensionless governing parameter $\xi=at$ 
not on $a$ and $t$ separately. That is,
we can write $\Pi=\phi(at)$. It implies that the solution for $C(a,t)$ must have the dynamic scaling form
\begin{equation}
\label{eq:scaling_form}
C(a,t)\sim t^\theta \phi(at),
\end{equation}
where $\phi(\xi)$ is known as the scaling function \cite{ref.family_Vicsek}.
The mass exponent $\theta$ is fixed by the conservation law. For instance, we can substitute Eq. (\ref{eq:scaling_form}) in 
Eq. (\ref{eq:conservation_law}) and we immediately find that $\theta=2$.

Now we attempt to verify Eq. (\ref{eq:scaling_form}) and find a solution for the scaling function $\phi(\xi)$. 
First, let us appreciate the fact that the three distinct curves in Fig. (\ref{fig2}) represent three distinct snapshots taken at 
three different times. It clearly reveals that for a given value of $a$ the numerical value of $C(a,t)$ is different 
for each different time. 
However, if the block area $a$ is measured using the inverse of time $t^{-1}$ as yardstick and $C(a,t)$ is measured
using $t^2$ as yardstick, then the numerical value of the corresponding dimensionless governed parameter $C(a,t)/t^2$ for a given
value of $at$  should coincide regardless of the size of the lattice or time $t$. 
That is, all the three distinct curves of Fig. (\ref{fig2})
should collapse onto one single universal curve if we plot $C(a,t)/t^2$ vs $at$ instead of plotting $C(a,t)$ vs $a$.
This is exactly what we have done in Fig. (\ref{fig3}) and found that the data points of all the three distinct curves of 
Fig. (\ref{fig2}) merge superbly onto a single universal curve which is essentially the universal scaling function $\phi(\xi)=C(a,t)/t^2$. 
Such data collapse of the distinct plots which represent three snapshots at different times 
implies that they are similar. However, as the same system at different times are similar we can regard
that the system is self-similar when it exhibits dynamic scaling.

It is clear from the log-linear plot of the scaling function that $\phi(\xi)$ decays exponentially in the large 
$\xi$-limit but it is certainly not exponential near the small $\xi$-limit. 
To find a better form for the scaling function we write the following trial solution 
\begin{equation}
 \phi(\xi)\sim \xi^{-\alpha}e^{-\xi}.
\end{equation}
We can find the $\alpha$ value by plotting $(at)^{\alpha}{{C(a,t)}\over{t^2}}$ versus $at$
in the $\log$-linear scale again and 
varying the $\alpha$ value till we get the best and longest strainght line. To this end, we find 
that $\alpha=1$ gives the most suitable straight line extending over the entire range of horizontal axis except in the vicinity
of the origin (see Fig. (\ref{fig3})). We therefore write the solution for the scaling function 
\begin{equation}
 \phi(\xi)\sim (at)^{-1}e^{-\xi}.
\end{equation}
It belongs to a different universality class than its one dimensional counterpart in which case,
$\phi(\xi)=e^{-\xi}$ \cite{ref.krapivsky}. The asymptotic solution for the area size distribution of the WPSL therefore is 
\begin{equation}
 C(a,t)\sim ta^{-1}e^{-at}.
\end{equation}
In fact, Krapivsky and Ben-Naim have shown analytically that the scaling function has two limiting behaviours which are consistent
with our simulation results shown in Fig. (\ref{fig3}). However, they predicted $\alpha=2$ for $\xi>>1$ but we found $\alpha=1$
instead.
\begin{figure}
\label{fig3}
\includegraphics[width=8.50cm,height=5.5cm,clip=true]{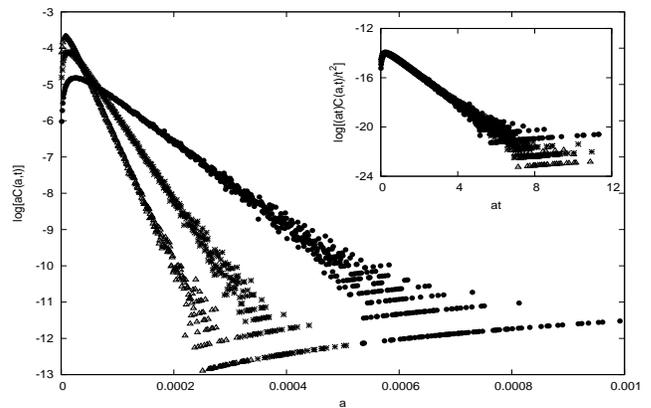}
\caption{Plots of $\log[aC(a,t)]$ is shown as function of $a$ for three different times. 
The resulting straight lines except near $a\rightarrow 0$ implies with slopes equal
to respective time the snapshots were taken implies that $C(a,t)\sim (at)^{-1} e^{-at}$.
\label{fig3}
}
\end{figure}

\begin{figure}
\includegraphics[width=8.50cm,height=5.5cm,clip=true]{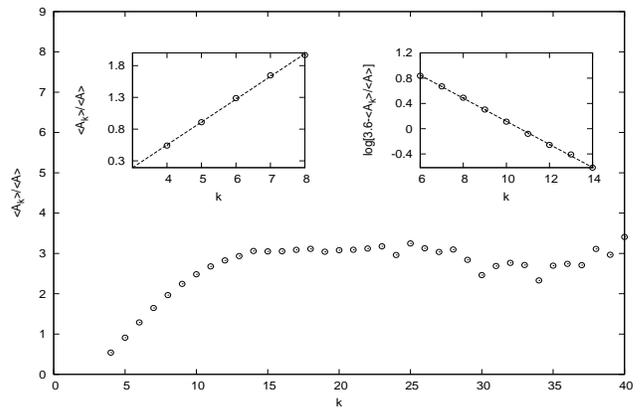}
\caption{Plots of $<A>_k$ vs $k$ to check if Lewis law is obeyed. The top left inset shows that the Lewis law is obeyed
for small $k$ value upto $k=8$. The top right inset, however, shows that beyond $k=8$ the $<A>_k$ grows exponentially and
saturates to its maximum value $<A>_k=3.5$.
\label{fig4}
}
\end{figure} 

\section{Lewis and  Aboav-Weaire laws}

A closer look at the snapshot of WPSL reveals that the neighbors of a block of given area can take up varying 
portions of its perimeter and hence can have neighbours more than the number of its sides.
Note that, whenever two blocks share a side even partially, they are considered as neighbors.
We therefore find it interesting to check if there exists a relation between the number of 
neighboring blocks and the corresponding block size or the area. 
In fact, it has been found in  several 
2D cellular mosaics empirically that the average area $<A>_k$  increases linearly with $k$ which is known as Lewis law. 
Whether this law is obeyed in the WPSL or not can be an interesting proposition. 
In Fig. (\ref{fig4}) we show how the mean area $<A>_k$ 
of the blocks which have exactly $k$ neighbours varies as a function of $k$. It clearly shows that the Lewis law is not
obeyed over the entire range of $k$ value. We can, however, identify three different regimes where relation
between $<A>_k$ and $k$ are signifiantly different. (i) The mean area $<A>_k$ increases 
linearly for small $k\leq 8$ revealing that the Lewis law is obyed. (ii) The mean area $<A>_k$ grows exponentially 
to a constant value for $8<k\leq 14$. (iii) Finally, the mean
area  $<A>_k$ stays constant for the entire range of $k>14$. These results are quite non-trivial
and they are in sharp contrast to the existing known results.

\begin{figure}
\includegraphics[width=8.50cm,height=5.5cm,clip=true]{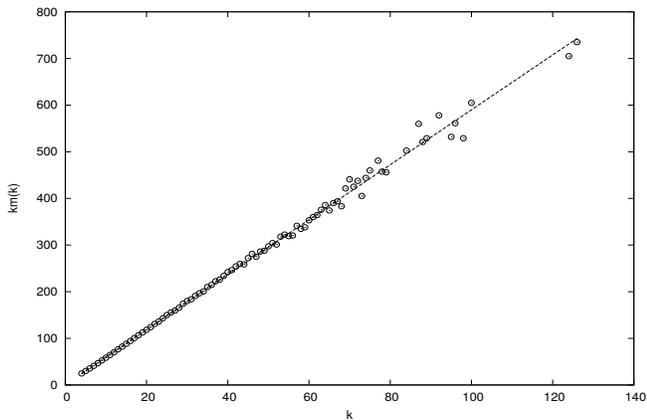}
\caption{We plot $km(k)$ vs $k$ and find a straight line passing through
origin with slope equal to $6$ revealing that the Aboav-Weaire law is violated.   
}
\label{fig5}
\end{figure}

\begin{figure}
\includegraphics[width=8.50cm,height=5.5cm,clip=true]{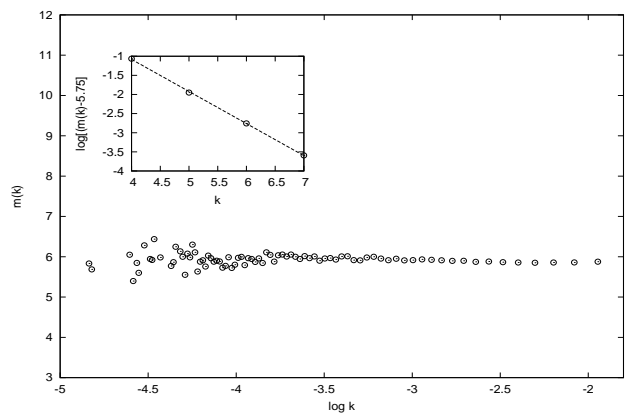}
\caption{Plots of $m(k)$ vs $k$ shows that for large $k$ it is almost constant. However, the plots of $\log(m(k)-5.75$
as a function $k$ in inset reveals that it reaches to constant $5.75$ exponentially. 
}
\label{fig6}
\end{figure}

The Aboav-Weaire law is yet another empirical law that has been found in several two-dimensional 
cellular structures. It was well studied particularly for soap froth. We find it worth checking
if this law is also obeyed in the case of WPSL. The Aboav-Weaire law given by Eq. (\ref{eq:aboav})
describe the short range topological correlations between 
$n$, the number of neighburs of a cell, and $m(n)$, the average number of neighbours of an adjacent block which
has $n$ neighbours. Upon multiplying Eq. (\ref{eq:aboav}) by $n$ on both sides we find an equation that
states that on average the sum of the number of sides $nm(n)$ of all the cells which are the neighbours of 
an $n$ sided cell is linear with $n$ having slope $a$ and intercept $b$. 
In the case of WPSL, we say a block has $k$ neighbours if all the $k$ block share a portion of their perimeter 
with it. We then plot $km(k)$ as a function of $k$ in Fig. (\ref{fig5}) and find a straight line with 
slope approximately equal to $a=6$. However, the key to the plot is that the intercept $b=0$. It implies that
the average number of neighbours of a neighbour of a block with $k$ neighbours is approximately $m(k)=6$ 
independent of $k$ which is shown in Fig. (\ref{fig6}).

\section{Summary}

We have studied a few interesting aspects of the weighted planar stochastic lattice (WPSL) which we have earlier shown to be a
multifractal and a complex scale-free network.
In this article, we primarily focused on its area size distribution function $C(a,t)$ and we have shown that it
exhibits dynamic scaling $C(a,t)\sim t^2\phi(at)$. We have verified it by using the idea of data-collapse. For this, 
we have shown that the distinct curves obtained by plotting $C(a,t)$ as a function of $a$ correponding to
different times collapses into a single universal curve if we plot $C(a,t)/t^2$ vs $x/t^{-1}$ instead. 
Such data-collapse means that the snapshots of the lattice at different times or of different sizes are similar 
in the same sense as two triangles are similar. One can show that the plots of area, say $A$, of right 
triangles as a function of one of the legs, say the opposite side $b$, will result in a  set of distinct curves for each different 
adjacent side $a$. However, if we plot
$S/c^2$ vs $b/c$ all these distinct curves of $S$ vs $b$ will collapse on to one single curve. Note that for a give value of
$b/c$ of a right triangle the numerical value of $S/c^2$ will coincide regardless of the size of the adjacent side $a$. We can thus
conclude that the collapse of the distincts curves means that the triangles are similar.
 
Besides, we have then investigated whether the Lewis and the  Aboav-Weaire laws are obeyed in the WPSL.
We found that the Lewis law is obeyed only for small $k<8$. It is clearly shown in Fig. (4) that $<A>_k$ is linear only
up to $k=8$. Beyond $k=8$ the Lewis law is violated. Instead, we found that $<A>_k$ increases 
exponentially and reaches to a constant value equal to $3$ beyond $k=14$. Finally, we investigated the 
Aboav-Weaire law and showed that it is violated for all $k$. Instead, we have found that the average number of 
neighbours of an adjacent block which has $k$ neighbours grows exponentially to a constant $5.75$. It implies that the ensemble
average of the mean number of neighbours of an adjacent block which has $k$ neighbours has $5.75$ neighbours. We
hope that the present study will further deepen our insight into various aspects of the weighted planar stochastic lattice.

\end{document}